\documentstyle[useAMS,psfig,epsf]{mn2e}

\newcommand{\re}{r_{\rm E}}

\newcommand{\amax}{A_{\rm max}}

\title{Improving the Prospects for Detecting Extrasolar Planets in Gravitational
Microlensing Events in 2002}
\author[I.A. Bond et al.]{I.A.~Bond,$^{1,2}$ 
F.~Abe,$^3$ 
R.J.~Dodd,$^{1,4}$
J.B.~Hearnshaw,$^2$ 
P.M.~Kilmartin,$^{1,2}$
\newauthor K.~Masuda,$^3$ 
Y.~Matsubara,$^3$ 
Y.~Muraki,$^3$ 
S.~Noda,$^3$ 
O.K.L.~Petterson,$^2$ 
\newauthor N.J.~Rattenbury,$^1$ 
M.~Reid,$^4$ 
To.~Saito,$^{}$ 
Y.~Saito,$^{3}$ 
T.~Sako,$^3$ 
J.~Skuljan,$^2$
\newauthor D.J.~Sullivan,$^4$ 
T.~Sumi,$^3$
S.~Wilkinson,$^4$ 
R.~Yamada,$^3$ 
T.~Yanagisawa,$^3$ 
\newauthor and P.C.M.~Yock$^1$
\\ 
$^1$Faculty of Science, University of Auckland, Auckland, New Zealand\\ 
$^2$Department of Physics and Astronomy, University of Canterbury, 
Christchurch, New Zealand\\ 
$^3$Solar-Terrestrial Environment Laboratory, Nagoya University, Nagoya 464, 
Japan\\ 
$^4$School of Chemical and Physical Sciences, Victoria University, 
Wellington, New Zealand\\ 
$^5$Tokyo Metropolitan College of Aeronautics, Tokyo 116, Japan\\ 
}

\date{Accepted .
      Received ;
      in original form }

\pagerange{\pageref{firstpage}--\pageref{lastpage}}
\pubyear{1994}

\begin{document}

\maketitle

\label{firstpage}

\begin{abstract}
Gravitational microlensing events of high magnification have been shown to
be promising targets for detecting extrasolar planets. However, only a few
events of high magnification have been found using conventional survey
techniques. Here we demonstrate that high magnification events can be 
readily found in microlensing surveys using a strategy that
combines high frequency
sampling of target fields with online difference imaging analysis.
We present 10 microlensing events with peak magnifications greater than 40
that were detected in real-time towards the Galactic Bulge during 2001 by MOA.
We show that
Earth mass planets can be detected in future events such as these through
intensive follow-up observations around the event peaks. 
We report this result with urgency as a similar number of such events are 
expected in 2002. 
\end{abstract}


\section{Introduction}

The importance of well-aligned or high magnification microlensing events for 
detecting planetary companions to the lens star was first pointed out by 
Griest \& Safizadeh (1998). They demonstrated that Jupiter-mass planets are 
readily detectable 
(if present) in events with maximum amplication, $A_{\rm max}$, as low as 10, 
and that Neptune-mass planets are detectable in events with $A_{\rm max} = 50$, 
if they are monitored intensely around their times of peak magnification. 
Unfortunately, the vast majority of the more than 1000 microlensing events 
that have now been detected by survey groups \cite{alc00,der01,udal00,bond01a} 
were of low magnification. We are aware of only one event with high 
magnification 
that received intensive observational coverage at its peak. This event, 
MACHO~98--BLG--35, reached $A_{\rm max}\sim80$. The observations 
yielded large exclusion regions for gas-giant planets surrounding the lens 
star, 
and also evidence for an Earth-mass planet near its Einstein ring 
\cite{rhie00,bond01b}. Two other high magnification events,
OGLE~00--BUL--12 and MACHO~99--LMC--2, received less intensive 
coverage yet still yielded large exclusion regions for gas-giant planets 
\cite{bond01b}.
An upper limit of the order of 30\% on the abundance of Jupiter-like planets
has also been obtained by the PLANET collaboration from a study of typical 
microlensing events \cite{alb01,gau02}. This study included the event 
MACHO~98-BLG-35 in which they excluded Jovian planets but were unable to 
draw any conclusions on the presence of terrestrial planets. This is not
inconsistent with the conclusions of Bond et al (2001b) since the PLANET
coverage of the peak of this event was less intensive and the data were 
analysed using a procedure that is generally less accurate.

Significant progress could clearly be made if one had a larger sample of high 
magnification events to work with. The purpose of this letter is to 
report the detection of 10 high magnification events thereby
demonstrating that high magnification events can be detected with high 
efficiency, and to urge follow-up observations of future events.

\section{Observations}

\begin{table*}
\centering
\caption{Details of high magnification events detected in 2001.}
\label{allevents}
\begin{tabular}{@{}rcrrccr@{}}

\multicolumn{1}{c}{Event} & 
\multicolumn{1}{c}{R.A.}   & 
\multicolumn{1}{c}{Dec.}   &
\multicolumn{1}{c}{$I_{\rm max}$} &
\multicolumn{1}{c}{$A_{\rm max}$} & 
\multicolumn{1}{c}{$t_{\rm E}$} & 
\multicolumn{1}{r}{$t_{\rm FWHM}$}\\

\multicolumn{1}{c}{} & 
\multicolumn{1}{c}{} & 
\multicolumn{1}{c}{} &
\multicolumn{1}{c}{} &
\multicolumn{1}{c}{} & 
\multicolumn{1}{c}{(days)} & 
\multicolumn{1}{r}{(hours)}\\
\\

2&
17 55 09.1&
$-$28 44 59.4&
11.6&
$42.6 \pm^{1.1}_{1.0}$ &
$38.2 \pm^{0.7}_{0.7}$&
74.6\\
5&
18 16 42.9&
$-$23 24 19.6&
14.3&
$200 \pm^{124}_{46}$&
$194 \pm^{17}_{46}$&
80.7\\
7&
18 08 58.8&
$-$27 36 11.9&
14.6&
$56 \pm^{17}_{11}$&
$22.0 \pm^{5.8}_{4.0}$&
32.7\\
16&
18 11 50.8&
$-$27 33 28.6&
$<$14.1&
$>$73&
$28.6 \pm^{3.8}_{9.4}$&
$<$32.6\\
20&
18 06 50.7 &
$-$27 15 13.3&
16.2&
$70 \pm^{326}_{40}$&
$14.9 \pm^{64.5}_{8.0}$&
17.7\\
32&
18 03 35.2&
$-$29 52 20.8 &
$<$14.1&
$>$96&
$17.6 \pm^{11.5}_{5.7}$&
$<$15.3\\
37&
17 55 23.4&
$-$28 56 44.2&
$<$14.5&
$>$186&
$>$23.8&
$\sim$10.6\\
41&
18 07 13.4&
$-$25 25 18.3&
$<$14.5&
$>$137&
$>$20.6&
$\sim$12.5\\
46&
17 57 48.9&
$-$29 36 35.5&
14.6&
$156 \pm^{320}_{82}$&
$16.9 \pm^{33.5}_{8.7}$&
9.1\\
50&
17 56 33.0&
$-$28 54 19.5&
$<$14.2&
$>$289&
$>$31.7&
$\sim$9.1\\

\end{tabular}
\end{table*}

\begin{figure*}
\centerline{\psfig{figure=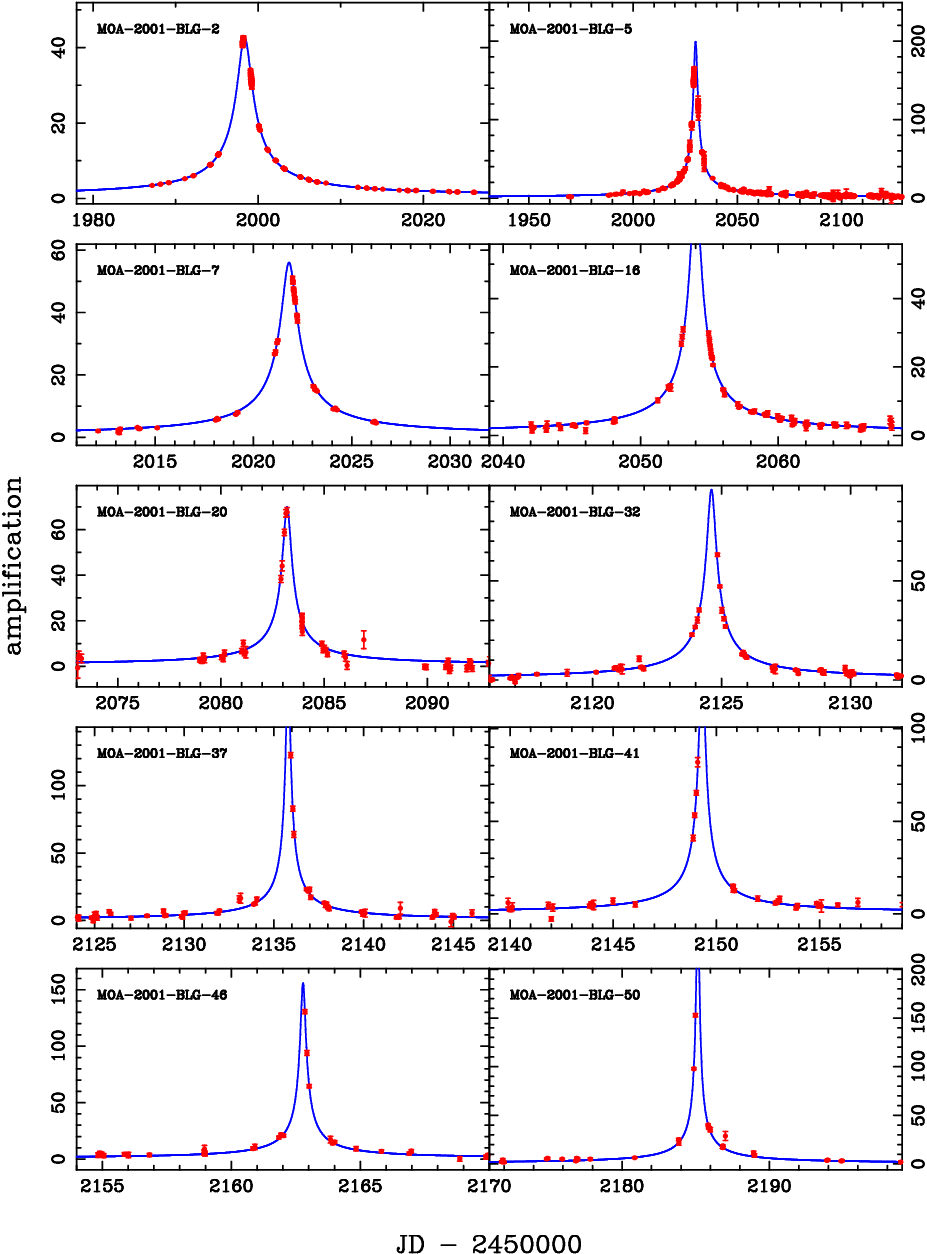,width=15cm}}
\caption{Light curves of high magnification microlensing events detected by
MOA during 2001. The photometric flux measurements have been converted to
amplifications using the fitted microlensing parameters given in Table 1.}
\label{fig.micro}
\end{figure*}

During 2000--2001 a campaign of observations was undertaken by the MOA 
collaboration with the aim of improving the detection rate of high 
magnification events. A 0.6m telescope at the Mt John Observatory in 
New Zealand (170$^\circ$ E, 44$^\circ$ S) with a mosaic camera comprised 
of three 2k $\times$ 4k thinned CCDs was used. An area 17 deg$^2$ towards 
the Galactic Bulge that is relatively unobscured by dust was monitored.

\begin{figure}
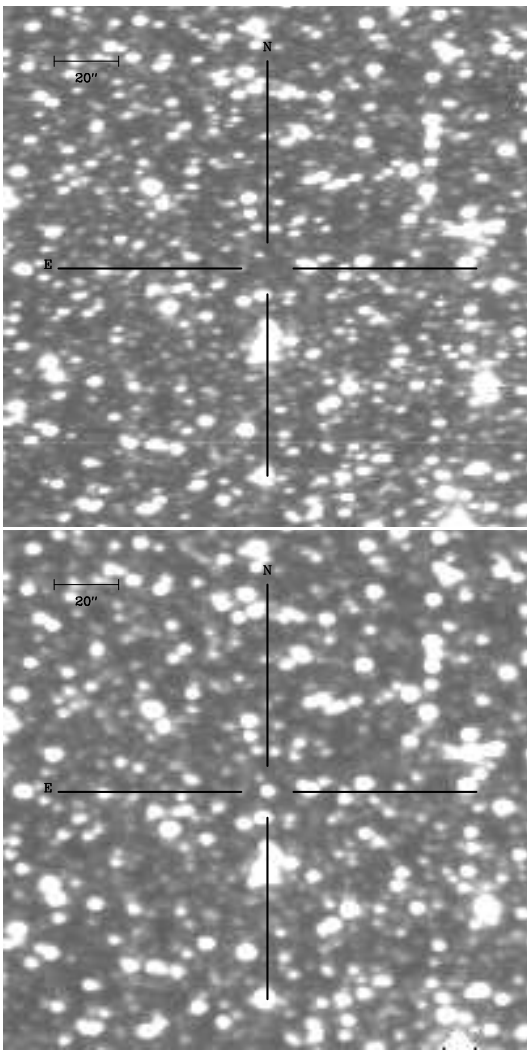

\centerline{\psfig{figure=mc017fig2a.ps,width=7cm}}
\centerline{\psfig{figure=mc017fig2b.ps,width=7cm}}
\caption{Sample images of event MOA~01--BLG--5 taken when the event
was at its baseline (top) and near its peak (bottom). The event stood
out clearly near its peak, but was invisible with the MOA telescope at its
baseline.}
\label{fig.frame}
\end{figure}

The MOA microlensing search procedure involves a combination of multiple
observations (up to six times) per night and real-time difference imaging 
to pick up microlensing events. 
Images taken during the pilot year of 2000 were used to build a database 
of variable stars 
and to detect some microlensing events \cite{bond01a}. The observations 
in 2001 were made primarily to search for microlensing events and to provide 
real-time 
alerts\footnote{http://www.phys.canterbury.ac.nz/$^\sim$physib/alert/alert.html}
to follow-up groups.
A total of 53 possible microlensing events were detected in real time in 2001,
of which 10 had $\amax > 40$. The details of these events are given in 
Table~\ref{allevents} and their light curves are shown in 
Fig.~\ref{fig.micro}. Most of the events
had $\amax \ge 100$ and all stood out clearly as shown, for example, in 
Fig.~\ref{fig.frame}. 

We determined the parameters $\amax$ and the the Einstein crossing time
$t_{\rm E}$ by fitting the standard single lens microlensing profile 
given by Paczynski (1986). The constraints on these parameters were determined
by a thorough examination over a range of values of $\amax$ and
$t_{\rm E}$. For some events only lower limits could be determined. The large
uncertainty in $\amax$ for some of the events in Table~\ref{allevents}
was mainly due to the less than complete coverage of the peak of their profiles.
The intrinsic faintnesses of the sources also contributed to the uncertainties.
With the exception of the first event, all events shown in Table~\ref{allevents}
had faint baseline intensities at or below the detection threshold.
However, since difference imaging was used, the photometry and subsequent
derivation of event parameters was unaffected by blending and thus free
from systematic effects that may result using conventional profile fitting
photometry techniques \cite{alc00,woz00,bond01a}

The MOA strategy of nightly multiple observations of survey fields is 
different from
that of earlier survey projects \cite{alc00,udal00,der01}. It is noteworthy
that most of the events shown in Table~\ref{allevents} and Fig.~\ref{fig.micro}
were observable only for a few days near their peaks. Events 32, 37, 41, 46, 
and 50 were especially rapid with most of the brightness changes ocurring in 
just one night. Such events had not been seen in previous surveys. If we
had adopted the strategy of just a single observation of the survey fields
per night, only events 2, 5, and possibly 7 would have been picked up. The
multiple nightly sampling strategy  would therefore appear to give 3--5
fold increase in the detection efficiency for rapid events.

Our use of difference imaging analysis has permitted the detection of events 
that would have otherwise 
been missed due to some combination of intrinsic faintness of the source 
and the degree of image crowding.  
Previous offline analyses using difference imaging have been carried
out on the image databases produced by MACHO \cite{alc00} and 
OGLE \cite{woz00} and have also shown that microlensing events with faint
source stars can be detected.
However, the sampling rate over the target fields in these surveys
was only of the order of once per night and events like the rapid ones 
presented appear to have been missed.

Our survey procedure also differs from previous ones in that the difference
imaging analysis is carried out on-line. This crucial feature enables events
to be alerted to other observatories in real-time for follow-up. Without this,
the potential for planet detection described in the following section
could not be realized. The real-time feature of our procedure was also
a factor in the scheduling of our survey fields.
For some of the events shown in Fig.~\ref{fig.micro}, the observation
frequency of those fields containing the events was increased following
the event alert. Unfortunately, bad weather and interruptions due to 
daylight prevented us from obtaining complete coverage of their peaks.

In general, the MOA survey procedure, which combines multiple observations 
of each of the survey fields nightly with on-line difference imaging analysis, 
evidently allowed us to tap into a huge 
reservoir of faint microlensing sources that had previously not been fully 
utilized. Our procedure enables detection of these events with small
telescopes during the brief time when they are highly magnified, i.e. when
the lens and the source are well-aligned. The fitted Einstein crossing times,
$t_{\rm E}$, for these events are not, however, unusually short. The crossing
time for an event depends upon the mass of the lens and the transverse velocity
of the lens with respect to source, and these parameters are not significantly
biased by our detection procedure. With the exception of 
event 5, which has an unusually long duration of $\sim 200$ 
days\footnote{Six similarly long timescale events with $t_{\rm E}>200$ 
days have been previously detected in the MACHO survey. These may be due to 
microlensing by stellar mass black holes \cite{benn01}}, 
all events had $t_{\rm E}$ 
in the range 10--40 days, consistent with the distribution found using the
conventional survey technique \cite{udal00}. Thus we see no evidence that
the MOA survey probes a different population of lens stars from that in 
previous surveys. However, it does probe a different population of source
stars. The data in Table 1 imply the source stars for all the 
events except the first are sun-like when allowance is made for reddening 
caused by dust \cite{sfd98}. These dwarf stars are ideal for planet hunting 
because source-size effects that tend to wash out planetary signals are 
minimised \cite{gs98}.

\section{Discussion}

It is evident from the above that microlensing events of high 
magnification with solar-type sources can be readily detected. 
Griest \& Safizadeh (1998) considered the detection of planets with masses 
ranging from that of Neptune up to that of Jupiter in microlensing events with 
$\amax$ up to 50. We have considered the detectability of lower mass 
planets in events with $\amax\ge50$ such as those reported here.
It is beyond the scope of this letter to consider in depth, the range of
sampling rates, detection criteria, telescopes, etc, that could be used
when monitoring high magnification events. We present here some
calculations as a benchmark as to what could be achievable in future events 
similar to those presented here. 

We simulated events with $\amax$ of 50, 100, and 200 that reached a peak I
magnitude of 15. We assumed they were monitored intensively during the
time interval $[-0.5t_{\rm FWHM},0.5t_{\rm FWHM}]$ where $t_{\rm FWHM}$
is the full-width at half-maximum of the microlensing light curve. This is
given by $t_{\rm FWHM}=3.5t_{\rm E}/\amax$ and is typically in the range 
10--30 hours. We assumed 300 measurements were made in this time interval 
with accuracy two times worse than the photon statistical limit achievable 
with a 1-m class telescope, consistent with accuracies attained using
the difference imaging technique. Full details of these simulations are
reported elsewhere (Rattenbury et al 2002, submitted to MNRAS). Simulated 
light curves corresponding to a 
range of planetary positions were generated and the quantity 
$\Delta\chi^2=\chi^2_{\rm single}-300$ was calculated for each light
curve. Here, $\chi^2_{\rm single}$ 
denotes the value derived by fitting the light curve for a single, 
planet-less lens to the simulated data for a lens with a planet. We adopted 
a detection criterion of $\Delta\chi^2 > 60$ corresponding to a $<$1\% 
probability for statistical fluctuations causing a planet-like signal.

\begin{figure}
\centerline{\psfig{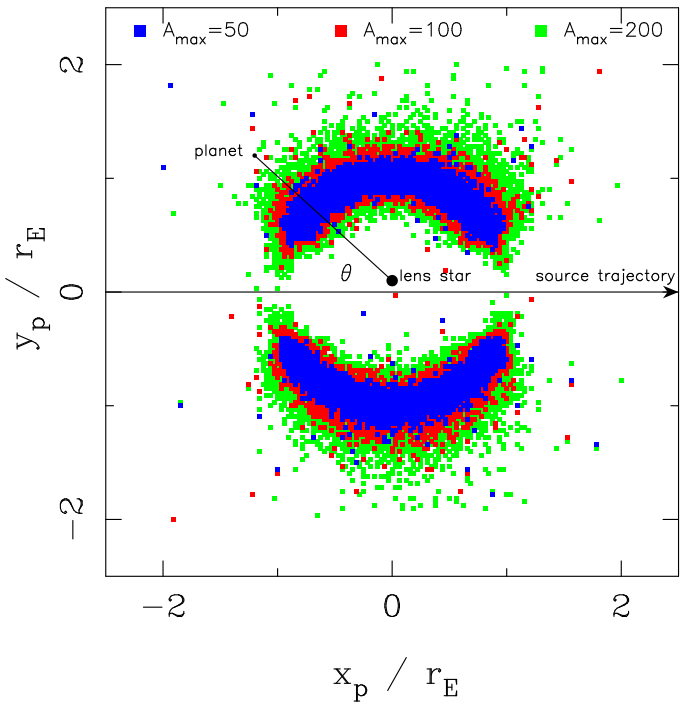}}
\caption{Detectability zones in the lens plane for an Earth-mass planet 
orbiting a $0.3M_\odot$ star in microlensing events with peak magnifications 
of 50, 100, and 200. The units for both axes are in $r_{\rm E}$ which is 
typically $\sim1.9$ AU. In this coordinate system, the lens star is located 
at position (0, $u_0$), where $u_0$ is the impact parameter, which is 
$\sim1/A_{\rm max}$. The vertical position of the lens star is exaggerated 
in the figure. If the projected position of the planet falls in the shaded 
regions at the time of the microlensing event, then it is detectable at the 
99\% confidence level if intensive photometric monitoring of the microlensing 
event is carried out during the time interval $t_{\rm FWHM}$. The angular 
diameter of the zone of detectability is $<$1 mas.}
\label{fig.sens}
\end{figure}

Fig.~\ref{fig.sens} shows the zones of detectability for an Earth-mass planet
under the above conditions. It is seen that Earth-mass planets can be detected 
with high efficiency if their orbital radii projected onto the lens plane
lies within the range 0.7--1.5$\re$ or $\sim$1.3--2.9 AU in high magnification
events that are monitored intensely during $t_{\rm FWHM}$. For larger planets 
heavier than Neptune, the zones of detectability extend far beyond the
region around the Einstein ring
especially when $\amax \ge 100$. Jupiter-mass planets are detectable 
almost anywhere in these events.

During the 2002 southern winter viewing season, we expect to detect 
around 10 high magnification microlensing events in real-time.
We propose that these events be monitored with a sampling as
dense as possible during the the time interval given by $t_{\rm FWHM}$
for the event. Given the rapid nature of many of these events, follow-up
observations will need to commence within 0--24 hours following the
event alert. For 
events with $t_{\rm FWHM}\sim24$ hours, a world-wide network of 1-m class 
telescopes is required to carry out the peak measurements. For more rapid 
events, fewer telescopes are required, but a commensurate increase in 
telescope aperture up to 2-m is required to achieve the sensitivity shown in
Fig.~\ref{fig.sens}. Difference imaging should be applied to all observations 
so as to achieve photometry with uncertainties
approaching that given by the photon noise and free from systematic effects
caused by blending of the source star with neighbouring stars. Looking further
ahead, continuous monitoring of the peaks of high magnification events
could be carried by follow-up observations from facilities in the Antarctic
or in space.

Follow-up measurements using ground or space based facilities with 
deep imaging capability would be required for all events in order to
determine to determine accurately the baseline intensity of the source
star. This information is required in order to determine the value of $\amax$
accurately and hence the planet:star mass ratio for any detected planet. In 
the case of non-detections, an accurate value of $\amax$ is required for 
the calculation of planetary exclusion regions. We note that these deep
follow-up observations are not time-critical and can be performed at any time
after the events. Further follow-up 
measurements a few years later with the Next Generation Space Telescope should 
enable the lens star in any event to be observed directly as it begins to 
diverge from the source star. This would enable the absolute 
value of the mass of a planet to be determined, and also its absolute 
instantaneous projected radius at the time of the microlensing event.

\section{Summary}

We have demonstrated that high magnification microlensing events 
can be readily detected in microlensing surveys with a strategy 
of high frequency sampling of survey fields and real-time difference 
imaging analysis. We have presented 10 events with $\amax\ge40$ that
were detected by the MOA microlensing survey during 2001. The purpose of this
letter is to bring this capability to the attention of the microlensing
community. We encourage intensive follow-up observations of future high
magnification events detected by MOA and similar experiments. 

Microlensing probes distant planets and is thus complementary to the
radial velocity and transit techniques.
Ultimately, a future space-based mission such as GEST \cite{benn02}, 
dedicated to observing thousands of microlensing events will be required in 
order to detect substantial numbers of low mass planets (including those 
with masses as low as that of Mars). In the meantime however, it would appear 
that intensive observations of high magnification events would provide the
best chances of detecting extra-solar planets in microlensing events 
observed from the ground.

\section*{Acknowledgments}

We thank P. Dobcsanyi, A. Gilmore, M. Honda, J.Jugaku, T. Nakamura, 
G. Nankivell, N. Rumsey, H. Sato, M. Sekiguchi, and Y. Watase for 
assistance, and the Marsden Fund of New Zealand and the Ministry of 
Education, Science, Sports and Culture of Japan for financial support.

\label{lastpage}


\begin{thebibliography}{}

\bibitem[\protect\citename{Albrow et al. }2001]{alb01} Albrow, M.D. et al. 
    (PLANET collaboration), 2001, ApJ, 556, L113 

\bibitem[\protect\citename{Alcock et al. }2000]{alc00} Alcock, C. et al. (MACHO collaboration) 2000,
    ApJ, 541, 734 
    
\bibitem[\protect\citename{Bennett et al. }2001]{benn01} Bennett, D.P., 
   et al. (MACHO and MPS collaborations) 2001, astro-ph/0109467

\bibitem[\protect\citename{Bennett et al. }2002]{benn02} Bennett, D.P., 
   et al. 2002, AAS meeting 199, \#9.05.

\bibitem[\protect\citename{Bond et al. }2001a]{bond01a} Bond, I.A. et al. 
   (MOA collaboration) 2001a, MNRAS, 327, 868
 
\bibitem[\protect\citename{Bond et al. }2001b]{bond01b} Bond, I.A. et al. 
   (MOA collaboration) 2001b, astro-ph/0102184

\bibitem[\protect\citename{Derue et al. }2001]{der01} Derue, F. et al. 
    (EROS collaboration) 2001, A\&A, 373, 126
    
\bibitem[\protect\citename{Gaudi et al. }2002]{gau02} Gaudi, B.S. et al. 
   (PLANET collaboration), 2002, ApJ, 566, in press

\bibitem[\protect\citename{Griest \& Safizadeh }1998]{gs98} Griest, K., 
   \& Safizadeh, N. 1998, ApJ, 500, 37

\bibitem[\protect\citename{Paczynski }1986]{pac86} Paczynski, B. 1986, 
   ApJ, 304, 1

\bibitem[\protect\citename{Rhie et al. }2000]{rhie00} Rhie, S.H. et al. 
   (MPS and MOA collaborations) 2000, ApJ, 533, 378

\bibitem[\protect\citename{Schlegel, Finkbeiner, \& Davis }1998]{sfd98} 
    Schlegel, D.J., Finkbeiner, D.P., \& Davis, M. 1998, ApJ, 500, 525
  
\bibitem[\protect\citename{Udalski et al. }2000]{udal00} Udalski, A. et al. 
    (OGLE collaboration) 2000, Acta. Astron. 50, 1

\bibitem[\protect\citename{Wambsganss }199?]{wamb} Wambsganss, J. 1997, MNRAS,
   284, 172 

\bibitem[\protect\citename{Wozniak }2000]{woz00} Wozniak, P.R. 2000, 
   Acta Astron., 50, 421 

\end{thebibliography}
\end{document}